\documentclass[structabstract]{aa}   % PRINTER FORMAT!!! el que uso siempre!
\usepackage{graphicx}
\usepackage[]{natbib}
\usepackage{txfonts}
\newcommand{\qub}{U-B \textit{vs} Q}

\newcommand{\Msun}{$\rm M_{\odot}$}
\newcommand{\Zsun}{$\rm Z_{\odot}$}

\newcommand{\Teff}{\mbox{$T_{\rm eff}$}}
\newcommand{\teff}{\mbox{$T_{\rm eff}$}}
\newcommand{\logg}{\mbox{$\log$~\textsl{g}}~}
\newcommand{\logq}{\mbox{$\log$~\textsl{Q}}~}

\newcommand{\kms}{km s$^{-1}$}

\newcommand{\halpha}{$\rm H_{\alpha}$}
\newcommand{\hbeta}{$\rm H_{\beta}$}
\newcommand{\hgamma}{$\rm H_{\gamma}$}
\newcommand{\hdelta}{$\rm H_{\delta}$}

\begin{document}
%
%   \title{Unveiling elusive O-stars in Local Group galaxies.}
   \title{The young stellar population of IC 1613. III.}

   \subtitle{New O-type stars unveiled by GTC-OSIRIS.
\thanks{
Based on observations made with the Gran Telescopio Canarias (GTC), 
instaled in the Spanish Observatorio del Roque de los Muchachos 
of the Instituto de Astrof\'{\i}sica de Canarias, in the island of La Palma.
Program ID GTC59-11B.}
       }
   \author{M. Garcia\inst{1,2}
          \and
          A. Herrero\inst{1,2}
          }

   \institute{Instituto de Astrof\'{i}sica de Canarias, C/ V\'{i}a L\'{a}ctea s/n E-38200 La Laguna, Tenerife, Spain.\\
              \email{mgg@iac.es}
         \and
             Departamento de Astrof\'{i}sica, Universidad de La Laguna, Avda. Astrof\'{i}sico
                Francisco S\'anchez, s/n, E-38071 La Laguna, Tenerife, Spain.
             }

   \date{Received july, 2012; accepted november 16th, 2012}

% \abstract{}{}{}{}{} 
% 5 {} token are mandatory
 
  \abstract
   { Very low metallicity massive stars are key to understand the reionization
epoch. Radiation-driven winds are one of the main agents of the evolution of massive stars,
and consequently an important ingredient of our models of the early-Universe.
%However, recent findings call our understanding of
%metal-poor massive star winds into question.
Recent findings hint that the winds of massive stars
with poorer metallicity than the SMC
may be stronger than predicted by theory.
Besides calling the paradigm of radiation driven winds into question, this result would
impact the calculated ionizing radiation and mechanical feedback of
massive stars, and the role these objects play at different stages of the Universe.
}
   {The field needs a systematic study of the winds of a large sample of very metal poor massive stars.
The sampling of spectral types is particularly poor in the very early types.
This paper's goal is to increase the list of known O-type stars in the dwarf irregular galaxy IC1613,
whose metallicity is smaller than the SMC's by roughly a factor 2.
}
   {
Using the reddening-free Q-parameter, evolutionary masses and GALEX photometry,
we built a list of very likely O-type stars.
We obtained low-resolution (R$\sim$1000) GTC-OSIRIS spectra for a fraction of them
and performed spectral classification, the only way to unequivocally
confirm candidate OB-stars.
}
   {We have discovered 8 new O-type stars in IC1613,
increasing the list of 7 known O-type stars in this galaxy by a factor of 2.
%  6 de Bresolin +  nuestra Of
The best quality spectra were analyzed
with the model atmosphere code FASTWIND to derive stellar parameters.
We present the first spectral type -- effective temperature scale
for O-stars beyond the SMC.
}
   {The target selection method is successful.
From the pre-selected list of 13 OB star candidates, we have found 
8 new O-stars, 4 early-B stars and provided a similar type for a 
formerly known early-O star. 
Further tests are needed but the presented procedure
can eventually make preliminar low resolution spectroscopy to confirm
candidates unnecessary.
%The presented selection procedure
%will save observing time by making preliminar low resolution spectroscopy to confirm
%candidates unnecessary. 
The derived effective temperature calibration for IC1613 is
about 1000K hotter than the scale at the SMC.
The analysis of an increased list of O-type stars 
will be crucial for the studies
of the winds and feedback of massive stars at all ages of the Universe.
}


        \authorrunning{M. Garcia \& A. Herrero}
        \titlerunning{New O-stars in IC1613}

   \maketitle
%
%________________________________________________________________

\section{Introduction}                   %___________________________________________________________

Massive stars are crucial to understand
the Universe because of their impact in many astrophysics fields.
Mighty stellar winds and ionizing radiation fields,
and a violent end as supernova,
disrupt their surrounding media and contaminate it
with the products of the nuclear reactions that feed these titans.
Our interest on very low metallicity massive stars
is growing rapidly because of their role in the early Universe.
In the primordial extremely metal-poor environment,
the formation of high mass stars was favoured \citep{BCL02} and these
%in turn, are intimately connected with
started
the re-ionization of the Universe.
Massive stars are suspected progenitors of 
long gamma ray bursts \citep[long-GRBs][]{Gal09,WH06},
and this connection may explain the highest
frequency of long-GRBs with higher redshift and 
decreasing metallicity \citep[][and references therein]{GRBs}.

%Radiation driven winds are one of the central pillars
%of the current paradigm of massive stars,
%as the wind removes mass from the star and changes the core conditions,
%regulates the duration of nuclear reactions and 
%ultimately decides the compact 
%object left by the SN explosion \citep{Gal09,WHW02}.
Blue massive stars (BMS) experience radiation driven winds,
powered by the scattering of photons in numerous UV transitions of metallic ions
and therefore strongly dependent on metallicity.
Radiation driven winds are one of the central pillars
of the current paradigm of massive stars,
as the wind removes mass from the star and changes the 
physical conditions at the stellar core.
The wind regulates, directly or indirectly the evolution of the star,
its ionizing radiation and mechanical feedback, the supernova (SN) explosion and
type of compact object left, and stellar yields \citep{Gal09,WHW02}.
%RDWs are included in the model atmospheres of blue massive
%stars --BMSs-- \citep{Pal05,WMB01,HM98} and the theoretical predictions
%are input into evolutionary codes \citep{MM00,Bal11}
%and in  models of galactic chemical evolution 
%\citep{HR10}.

Theory predicts a strong correlation between 
the momentum carried by the wind and
the luminosity of the star
and metallicity: the wind-momentum luminosity
relation \citep[WLR,][]{Kal95}.
This relation and its metallicity dependence
have been thouroughly characterized 
both by theoreticians \citep[e.g.][]{VKL01}
and observations \citep{Mal07}
from the Milky Way (MW) down to the metallicity of the
Small Magellanic Cloud (SMC).
Very low metallicity BMS are expected to experience 
weaker winds than SMC stars and much weaker winds than MW stars. However, 
some recent results are in marked contrast.

We have found a resolved Luminous Blue Variable star (LBV)
of very low  metallicity \citep[$\rm \sim 0.2 Z_{\odot}$,][]{Hal10}
with strong optical P~Cygni profiles;
an analogous example exists in NGC2366 \citep{DCS01}
and similar unresolved cases may exist in more distant galaxies \citep{pustilnik08,izotov11}.
%The metal-poor dwarf irregular IC1613 hosts one of the few WO known in the Local Group \citep[WO3,][]{KB95},
%an extreme type of Wolf-Rayet star stripped off its outer envelope by the stellar wind.
\citet{Tal11} reported 6 stars with stronger wind momentum than expected
at the poor metallicity of their host galaxies ($\rm \sim 1/7 \, Z_{\odot}$)  %IC1613, WLM and NGC3109 
from X-Shooter spectroscopic analyses, although error bars are too large for results to be determinant.
Our analysis of an Of star in IC\,1613 \citep{Hal12} concluded that 
the star may have a strong wind or, alternatively,
a slower than expected wind acceleration.
\citet{L12} argues that the neglect of wind clumping may explain these findings,
but also suggests that the discrepancy may be caused by an additional wind-driving mechanism
operating only at certain metallicities, temperatures and luminosities,
or negligible in the Galactic counterparts.
\citet{Hal12}'s and \citet{Tal11}'s examples, if confirmed by the detailed study of a large sample of objects,
pose a challenge to the standard theory of radiation line-driven winds, as there are few metals to drive the wind.
%and would imply a change in the calculated wind strength of blue massive stars but also in their emergent UV
%ionizing radiation.
Yet, this might explain why
long-GRBs (typically associated with type Ic SN, \citealt{WH06})
are mostly found in metal-poor environments \citep{Mal08,Lal10}
but require a strong wind to remove the H and He envelope in the pre-SN stages.

Besides the wind mechanical energy,
the ionizing radiation emitted by massive stars
is a chief interface of the pre-SN stages with the interstellar medium.
The production of ionizing photons can be estimated to first
order from the effective temperature (\teff), making \teff~\textit{vs} spectral type calibrations
a very useful tool.
While the temperature scale for the Milky Way and the Magellanic Clouds
have been addressed by a number of works \citep[e.g.][]{GB04,Mal05b,Mal09},
no characterization for lower metallicity environments exists.

In order to study very metal-poor BMSs
we need to reach out into the Local Group, beyond the Magellanic Clouds \citep{NUVA}.
IC1613 is the closest Local Group galaxy (DM=24.27 \citet{Dal01}; E(B-V)=0.02 \citet{LFM93})
with on-going star formation and poorer metallicity than the SMC: 
$\rm \sim 0.13 Z_{\odot}$ from B-supergiants \citep{Fal07},
$\rm \sim 0.05 Z_{\odot}$  from nebular studies  \citep{Fal07,PBT88,T80,DK82}.
\citet{Fal07} published a catalog of about 40 OB-stars in this galaxy,
found from VLT-FORS2 MOS observations, but it only lists 6 O-types
(4 of them were observed and analysed by \citet{Tal11}).
These, together with the Of star analyzed in \citet{Hal12} makes a total of 7 O-type stars known
in IC1613, insufficient for a statistical characterization of
the WLR at very low metallicities.

It is necessary to extend the sample of known O-stars in IC1613,
but so far low-resolution spectroscopy is needed to unequivocally confirm O-candidates.
In this paper we present new OB-stars in IC1613
found from low-resolution spectra
taken with the Gran Telescopio Canarias (GTC).
The targets were selected using a set of photometry-based criteria
which works well for O-stars
as endorsed by the derived spectral types.
This work is part of our study of the BMS population of IC1613
and uses the photometric catalog 
and study of OB~associations published in
\citet[][hereafter GHV09]{GHV09}
and
\citet[][hereafter GHC10]{Gal10}.
All stellar identification numbers 
refer to the GHV09 catalog,
unless otherwise specified.

We present the target selection procedure optimized for O-type stars 
in Sect.~\ref{S:selec}.
The observations and data reduction are detailed
in Sect.~\ref{S:obs}.
In Sect.~\ref{S:SpT} we provide spectral types for the targets,
and in Sect.~\ref{S:FW} we derive their stellar parameters.
Finally, our conlusions are provided in Sect.\ref{S:con}.

%____________________________________________________________________________________________________

\section{Target Selection}               %___________________________________________________________
\label{S:selec}

% esta esta cogida directamente de
%    /scratch/scratch2/mgg/proposals/GTC/2011B/GTC_IC1613/PHASEII/targets_GTC59_11B.sky
% que contiene todas mis target stars!!

\begin{table*}
\caption{Coordinates, photometric data and association membership for the sample stars,
from GHV09. Stars not belonging to any OB association are marked [-1] in colum ASSOC.
The spectral types provided in column SpT were derived in this work.}           
\label{T:assoc}      
\centering        
\begin{tabular}{r r r r r r r l l}     
\hline\hline  
ID    & RA[deg]   & DEC[deg]      &ASSOC&     V &   B-V  &     Q &          SpT  & Notes                        \\
      & J2000.0   & J2000.0       &     &       &        &       &               &                              \\
\hline
64066 &	16.258634 &	 2.157802 & 147	& 19.03	&  -0.21 & -0.97 &   O3~III((f)) &                              \\
65426 &	16.262785 &      2.167927 & 162	& 19.62	&  -0.20 & -0.91 &        O6~III & offslit, O5-6~V \citep{Fal07}\\    
69476 &	16.277063 &      2.158948 & 185	& 18.89	&  -0.22 & -0.97 &      O6.5~III &                              \\
36611 &	16.186651 &      2.109095 & 44	& 19.46	&  -0.13 & -0.91 &      O7~III-V & multiple?                    \\     % multiple tanto x visual insp. como x stellar substruct.
75661 &	16.307566 &      2.139771 & 197	& 19.79	&   0.18 & -1.07 &        O8~III &                              \\    
67684 &	16.270335 &      2.159043 & 175	& 19.02	&  -0.21 & -0.91 &        O8.5~I & composite?                   \\     % esto es pq He-lines parecen embebidas en broad bands.
60782 &	16.249088 &      2.153501 & 127 & 19.61 &  -0.22 & -0.91 &      O9.5~III & blend?                       \\     % esta es la unica que no estaba en targets_GTC59_11B.sky
61331 &	16.250685 &      2.153629 & 127	& 19.14	&  -0.21 & -0.85 &       O9.7~II & elongated shape but single detection in GHV09 \\    
71708 &	16.286754 &      2.152986 & 192	& 19.78	&  -0.02 & -1.04 & lateO~III+neb &                              \\    
60269 &	16.247649 &      2.153412 & 127 & 20.49 &  -0.29 & -0.82 &        B0.5~I & offslit                      \\
60882 &	16.249384 &      2.096583 & 135	& 19.44	&  -0.14 & -0.84 &    B0.5~I-III &                              \\                                                        
27381 &	16.163216 &      2.146994 & -1	& 18.42	&  -0.07 & -0.83 &      B1-1.5~I &                              \\   
35071 &	16.182901 &      2.084464 & -1	& 18.78	&   0.65 & -0.99 &      B2.5~III & V39-like colors. Double?     \\                                                      
44736 &	16.206282 &      2.106804 & 56	& 19.89	&  -0.12 & -0.81 &               & offslit                      \\
67063 &	16.268118 &      2.163212 & 176	& 19.65	&  -0.04 & -0.96 &               &                              \\
68456 &	16.273197 &      2.153952 & 175	& 21.67	&   0.19 & -1.35 &               & WR-candidate                 \\  
\hline
\end{tabular}
\end{table*}
  %\label{T:assoc}

O-type stars cannot be chosen solely from optical photometry,
since their colors are very similar to those of the slightly cooler B-types.
So far spectral types are needed to unequivocally spot out O-stars,
but this may be costly in observing time since 
they are often obscured by gas and dust and quite faint.
We therefore need to device a method to produce a sound list
of very likely O-type stars,
and additional criteria (besides photometric colors) are needed to this end.

\begin{figure}
\centering
\includegraphics[width=0.45\textwidth]{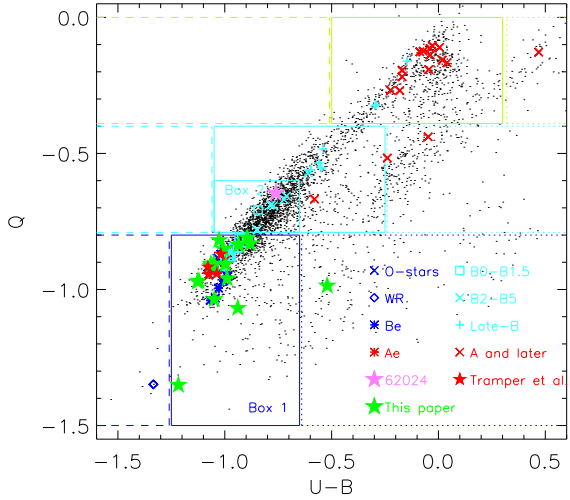}
   \caption{IC1613's \qub~ diagram. Black dots mark catalog stars with high quality photometry
%\citep[see][]{GHV09}. 
from GHV09.
Other colors and symbols, except for filled stars, mark the position
of stars with known spectral types from \citet{Fal07}.
O- and early-B- stars concentrate in boxes
1 and 2.
The pink star represents 62024, an Of star analyzed by \citet{Hal12}.
Red stars represent the sample of \citet{Tal11} in IC1613.
Green stars mark this paper targets.
           }
      \label{F:QUB}
\end{figure}

\begin{figure}
\centering
\includegraphics[width=0.45\textwidth]{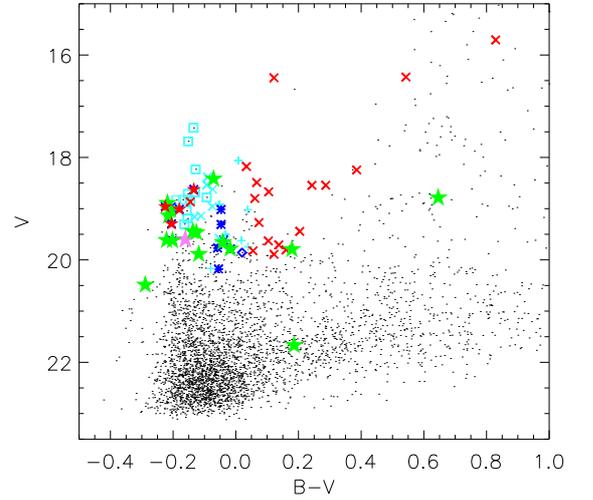}
   \caption{IC1613's color-magnitude diagram.
Colors and symbols as in Fig.~\ref{F:QUB}.
OB-stars are dispersed in the color-magnitude diagram,
and do not concentrate on the galaxy's blue-plume.
In sight of this diagram three targets are remarkably reddened,
and would have been discarded with classical $B-V$~ color cuts.
Note also that the brightest stars of the blue-plume
are early-B supergiants.
           }
      \label{F:CMD}
\end{figure}

GHV09 showed that OB-stars are found in a particular
locus of the \qub~ diagram (see Fig.~\ref{F:QUB} and GHV09-Fig.~5), 
Q being the reddening-free Q pseudo-color $Q \, = \, U-B \, - \, 0.72 \times (B-V)$.
The explanation is that Q 
increases monotonically towards later spectral types in the interval Q$\in$[-1.0,-0.4], 
corresponding to O3-A0 types (see e.g. the calibration of \citet{F70}, and GHC10).
While Q is an indicator of spectral type,
U-B holds the information of whether the star is reddened or not. 
Q is the primary target selection criterion:
starting from a catalog with small photometric errors ($<$0.05mag)
to minimize the impact on Q,
we choose "blue-Q" stars with $\rm Q < -0.8$.
An additional advantage of choosing stars mainly from their Q pseudo-color
instead of classical B-V cuts, is that locally reddened targets 
are not discarded (see Fig.~\ref{F:CMD}).

Yet, in order to separate  O and B stars (see GHV09)
further information is needed.
%Targets were chosen from our catalogs and
%composite RGB images made with archival 
%INT-WFC \halpha~ and V-band images, and GALEX-FUV channel (see Fig~\ref{F:RGB})
%using the Aladin tools \citep{aladin}.
From IC1613's ``blue-Q'' stars, we chose those
with the largest evolutionary mass ($>$ 25 \Msun, derived by GHC10)
that matched stellar-like sources in GALEX-FUV images.
We allowed for magnitudes as faint as $V=19.8$~ as long as the GALEX detection
is clear.
%; brighter stars are probably B-supergiants.
%The main slit targets were chosen outside large HII structures
%whenever possible.

%This procedure for target selection is quite successful, as we will discuss below,
%but not free of caveats.
The procedure is not free of caveats.
From its definition it follows that 
the Q-parameter depends on the adopted reddening law and 
is subject to larger photometric errors than individual filters.
Besides, some O stars exhibit $\rm Q > -0.8$~ and would not be included in the current selection.
Nonetheless, our selection criteria make the hunt for O-type stars more effective.

%We submitted a low-budget proposal (5h granted) to look for new O-stars in IC~1613.
%Candidate stars, chosen from their reddening-free Q color
%(Q=U-B - 0.72*(B-V)), GALEX detection and evolutionary mass, must be first
%confirmed with spectral classification prior to investing 
%previous telescope time to obtain good quality spectra apt for quantitative analysis.
%Our proposal was designed to obtain low resolution spectra (R2000B, 1.2\arcsec slit)
%of 2 candidates at once per slit per hour in order to obtain
%spectral types and identify the most interesting candidates for subsequent follow-up.

\begin{figure}
\centering
\includegraphics[width=0.45\textwidth]{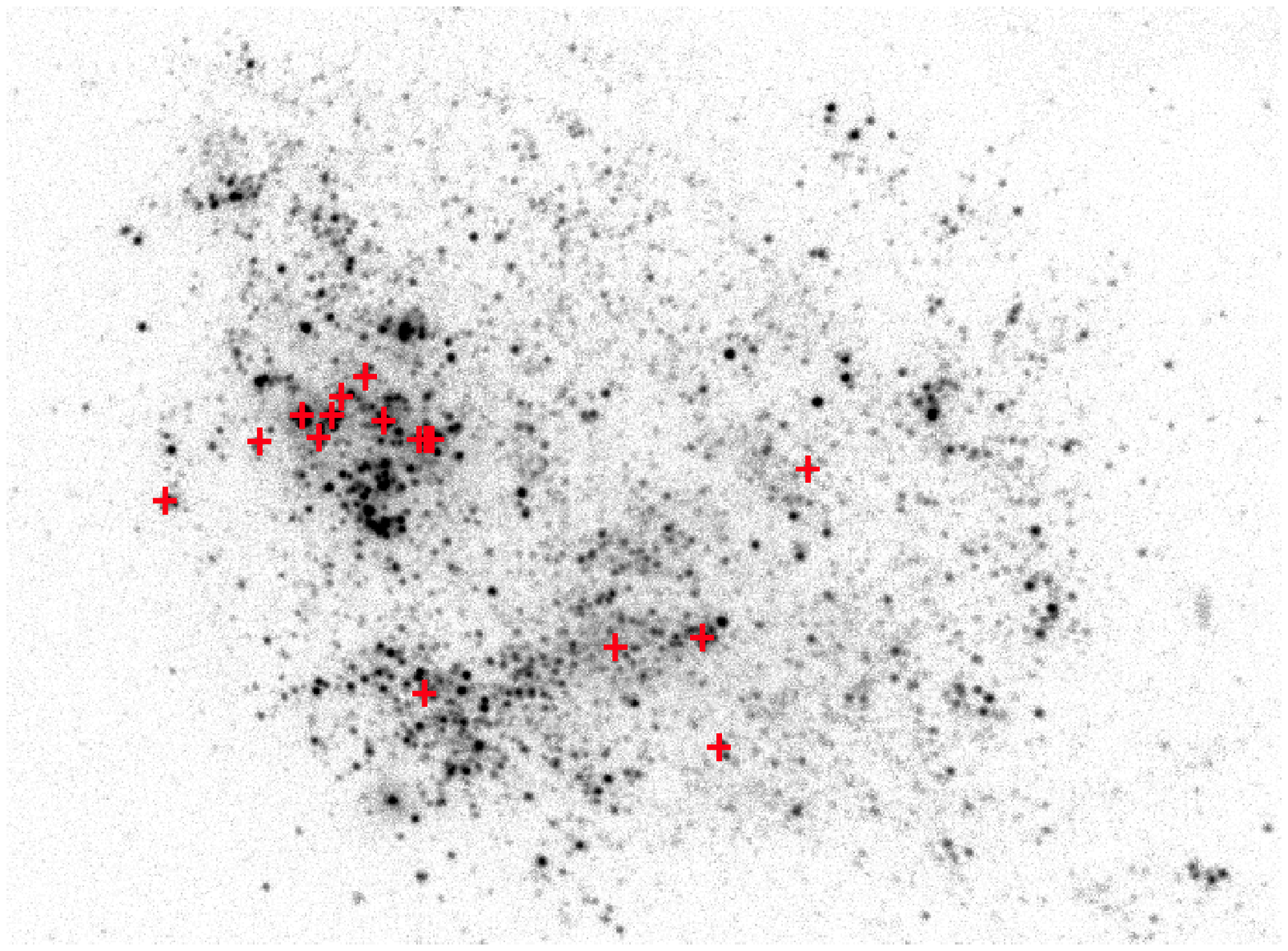}
   \caption{IC1613, GALEX's FUV-channel. North is up and East is left.
Plus symbols mark the position of the sample stars,
chosen with the following criteria:
Blue Q-parameter ($Q < -0.8$), V-magnitude ($V < 19.8$), intense emission
in GALEX images, large evolutionary mass ($M > 25$\Msun)
and no \halpha~ emission whenever possible.
%In this RGB image, our candidate OB stars are blue (hence intense FUV radiation)
%and without red around (hence no \halpha~ emission).
           }
      \label{F:RGB}
\end{figure}

The list of candidate OB-stars and their photometry is provided in Table~\ref{T:assoc}.
We also included in the spectroscopic follow-up two objects with colors
similar to the known WO and LBV-candidate stars in IC1613 respectively (see GHV09),
one previously known O~ star (65426, O5-6~V, \citet{Fal07}),
and two additional blue but fainter stars that serendipitously falled in-slit.
Their location in IC1613 is shown in Fig.~\ref{F:RGB}.
Figs.~\ref{F:QUB}~and~\ref{F:CMD} illustrate their position in the
\qub~ and color-magnitude diagrams.
Finding charts are provided in Appendix~\ref{S:f-charts}.

%____________________________________________________________________________________________________

\section{Observations and data reduction}                  %___________________________________________________________
\label{S:obs}

The observations were carried out with 
the Optical System for Imaging and low
Resolution Integrated Spectroscopy (OSIRIS)
at the 10m-telescope GTC.
The program was granted 5 hours under proposal GTC59-11B (P.I. M.Garcia),
in service mode.
The observations consisted on long-slit spectroscopy taken with the R2000B VPH
and 1.2$\arcsec$ slits.
The spectra cover the $\sim$4000-5500 \AA~ range
with resolving power $\rm R \sim 1000$, suitable to perform spectroscopic classification.
The spectra are well oversampled (about 5 pixels)
and can be rebinned to improve the signal-to-noise ratio (SNR).

The slits were oriented in specific angles
to include several targets.
%(the 16 program stars were observed in only 5 slits).
%Each slit was observed during 1 hour long Observing Block (OB),
%broken into two exposures of 1249s each, to enable cosmic ray subtraction.
One 1-hour long Observing Block (OB) was devoted to each slit.
The observing conditions are compiled in Table~\ref{T:log}.

% AQUI ESTAN TODAS LAS ESTRELLAS QUE YO METI EN LAS RENDIJAS
  %  60782 entro de casualidad

\begin{table*}
\caption{Observation log: slit position angles (PA), included targets, date of observations, 
seeing and transparency conditions, fraction of illuminated moon and airmass.} 
\label{T:log}      % is used to refer this table in the text
\centering                          % used for centering table
\begin{tabular}{l l l l l l l l l}        % centered columns (4 columns)
\hline\hline               
  Slit   &  PA   &  Stars                     & Date       & Seeing     & Trans.& Moon  &  Airmass \\    % table heading 
         & [deg] &                            &            & [$\arcsec$]&       & [FLI] &          \\
 \hline                                                                                 
 slit\_a & 85.6  & 61331, 27381, 60269, 60782 & 10-02-2011 & 0.9        & Spect.& 0.40  &  1.15    \\
                                                                                        
 slit\_b & -57.8 & 69476, 71708, 75661, 65426 & 10-24-2011 & 1.2        & Clear & 0.05  &  1.12    \\    % airmass calculated as the average of both exp.
                                                                                        
 slit\_c &  45.85& 64066, 35071, 44736        & 09-17-2011 & 0.9        & Clear & 0.70  &  1.24    \\
                                                                                        
 slit\_d & -28.6 & 67684, 67063, 68456        & 09-16-2011 & 0.9        & Clear & 0.80  &  1.44    \\
                                                                                        
 slit\_e & -78.8 & 36611, 60882               & 09-17-2011 & 1.0        & Clear & 0.70  &  1.66\tablefootmark{a}    \\

\hline                     
\end{tabular}

\tablefoottext{a}{Morning twilight started few minutes before completing
the second exposure}

% el seeing fue medido por MGG

\end{table*}
    % \label{T:log}

%____________________________________________________________________________________________________

\subsection{Data reduction}                      %___________________________________________________________

%\begin{table}
%\caption{Runs para cada OB}             % title of Table
%\label{T:OB}                         % is used to refer this table in the text
%\centering                          % used for centering table
%\begin{tabular}{l l l l}        % centered columns (4 columns)
%\hline\hline                 % inserts double horizontal lines
%        &   run\#1   &   run\#2 &  run\#3      \\    % table heading 
%\hline                        % inserts single horizontal line
%OB0001  &   No vale  &   OK     &              \\
%OB0002  &   No vale  &   OK     &              \\
%OB0003  &   No vale  &   Dudosa &  planned     \\
%OB0004  &   No vale  &          &  planned     \\
%OB0005  &   No vale  &          &  planned     \\
%\hline                                   %inserts single line
%\end{tabular}
%\end{table}

% Esto viene de AAAstrategia.txt :

Reduction was performed according to standard IRAF\footnote{
IRAF is distributed by the National Optical Astronomy Observatory, which is operated by the 
Association of Universities for Research in Astronomy (AURA) under cooperative agreement with the National Science Foundation
}
procedures.
Each slit OB was split into two 1249s exposures, to enable cosmic ray removal.
%Two exposures of 1249s were taken per slit, to enable cosmic ray removal.
The exposures were coadded prior to any reduction procedure (case-A),
with the \textit{imcombine} routine and the \textit{crreject} mechanism for cosmic-ray rejection.
We checked against displacement between consecutive exposures
and found shifts of up to 7 pixels in extreme cases (slightly larger than 1 resolution element)
in the spectral direction.
For this reason we also reduced the individual exposures of all stars separately,
and coadded the two spectra at the end of the reduction (case-B),
to avoid problems with wavelength shifts.
We compared the resulting spectra from case-A and case-B reduction 
and found undetectable differences,
with a better cosmic-ray extraction in case-A, which we adopt.

The \textit{ccdproc} routine was used to trim the images and remove
vignetted areas of the CCD, and also for bias subtraction and flat-field correction.
The bias correction was performed using the overscan region.
Each spectrum was flat-field corrected with
the normalized attached daytime flats for that OB.
%Bias correction was performed using the overscan region, since
%the count-level of observed structures amounts to less than 1\% of the total counts.
%Each spectrum was flat-field corrected with
%the attached daytime flats for that OB.
%They were normalized prior to flat-field correction with the \textit{response} task.
After checking against shifts between 
the master arcs provided by the observatory and the attached arcs,
the former were used for wavelength calibration because of their increased SNR.
%\textbf{OSIRIS's projection of spectral lines on the detector
%is stable. Master-arcs provided by the observatory with increased
%SNR over the attached OB arcs were used for the wavelength calibration,
%after checking against shifts between master and attached arcs.}
A 2-D wavelength calibration solution was obtained
and assigned to the data using iraf routines
\textit{identify, reidentify, fitcoords} and \textit{transform}.
% esto es lo que da trasnform:
%Spectral resolution for chip-1 is 0.84\AA~ and 0.86\AA~ for chip-2;
%for stars close to the right edge of chip-1 there is difuse light
%and the last arc line can be hardly seen --> wavelength calibration is
%not reliable from 5500\AA~ on.

We note that the arc lamp image displays difuse light on the right part of chip-1
at $\lambda \gtrsim$ 5500\AA, and the red wavelength calibration 
for stars whose spectra laid in this region of the CCD is not reliable.
The GTC observatory also reported intermitent technical problems with OSIRIS's
active collimator during the dates of the observations, which may affect the stability of the 
wavelength calibration.
Both factors could explain the wavelength calibration problem reported in Sect.~\ref{S:FW}.
However, neither of them hamper the identification of spectral lines
or the derivation of stellar parameters,
hence have no negative impact on this paper goals.
%These issues do not hamper the identification of spectral lines,
%hence have no negative impact on this paper goals.

The \textit{apall} task was used for target extraction and 
background subtraction.
Background subtraction was complicated since a fraction of the
targets were located in crowded regions,
and some of them also experienced nebular contamination.
%Background subtraction was complicated since a fraction of the
%targets were located in crowded regions,
%and some of them also experienced nebular contamination
%despite our efforts during target selection.
The fit to background levels was performed at the physical line of the image
equivalent to 4500\AA, free from nebular transitions.
If two stars were very close together, their apall aperture was 
chosen to minimize contamination from the neighbor. 
Some targets exhibit a pedestal in the transversal cut,
which could be caused by either nebular contamination or crowding.
We checked \halpha~ images to evaluate
the first option and, in the event of \halpha~ detection,
the sky aperture was chosen close to the star, at the pedestal.
If there was no \halpha~ detection, then we interpreted that the plateau had stellar origin,
and the sky contribution was chosen far from the pedestal.

Finally, the R2000B VPH of GTC-OSIRIS  experiences a ghost image, with negligible number
of counts in the science or arc images.
However, a non-negligible structure remains in the flat-field after normalization, 
which is transmitted to the science image 
after the flat-field correction.
The ghost position varies between different  flat-field images,
but we have made a conservative estimate of the wavelength ranges
that may be affected:
       4738 - 4778\AA~ (chip1)
and                       
       4736 - 4784\AA~ (chip2).

\begin{figure*}
\centering
   \includegraphics[width=\textwidth]{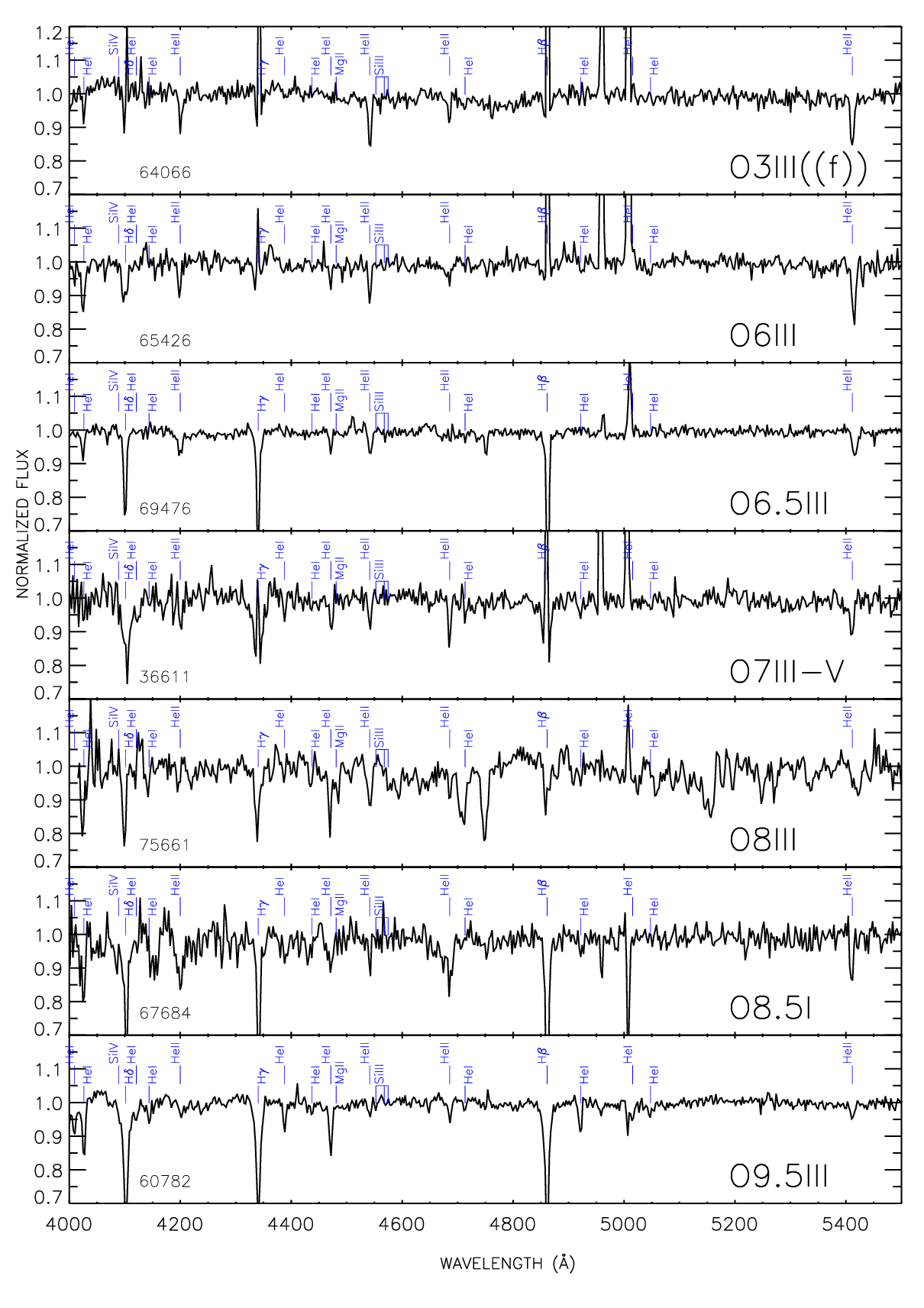}
   \caption{GTC-OSIRIS spectra of OB-type stars in IC1613.
The spectra have been rebinned to 3 pixels and corrected 
by each star's radial velocity except for 71708, % and 67063,
for which we used IC1613's systemic velocity -234 $km \, s^{-1}$ \citep{Lal93}.
%\sc{H}{i}, \sc{He}{i} and \sc{He}{ii} spectral transitions are marked as reference. 
The evolution of HeII4541/HeI4471 is clearly seen.
}
   \label{F:all0}%
\end{figure*}

\begin{figure*}
\centering
   \includegraphics[width=\textwidth]{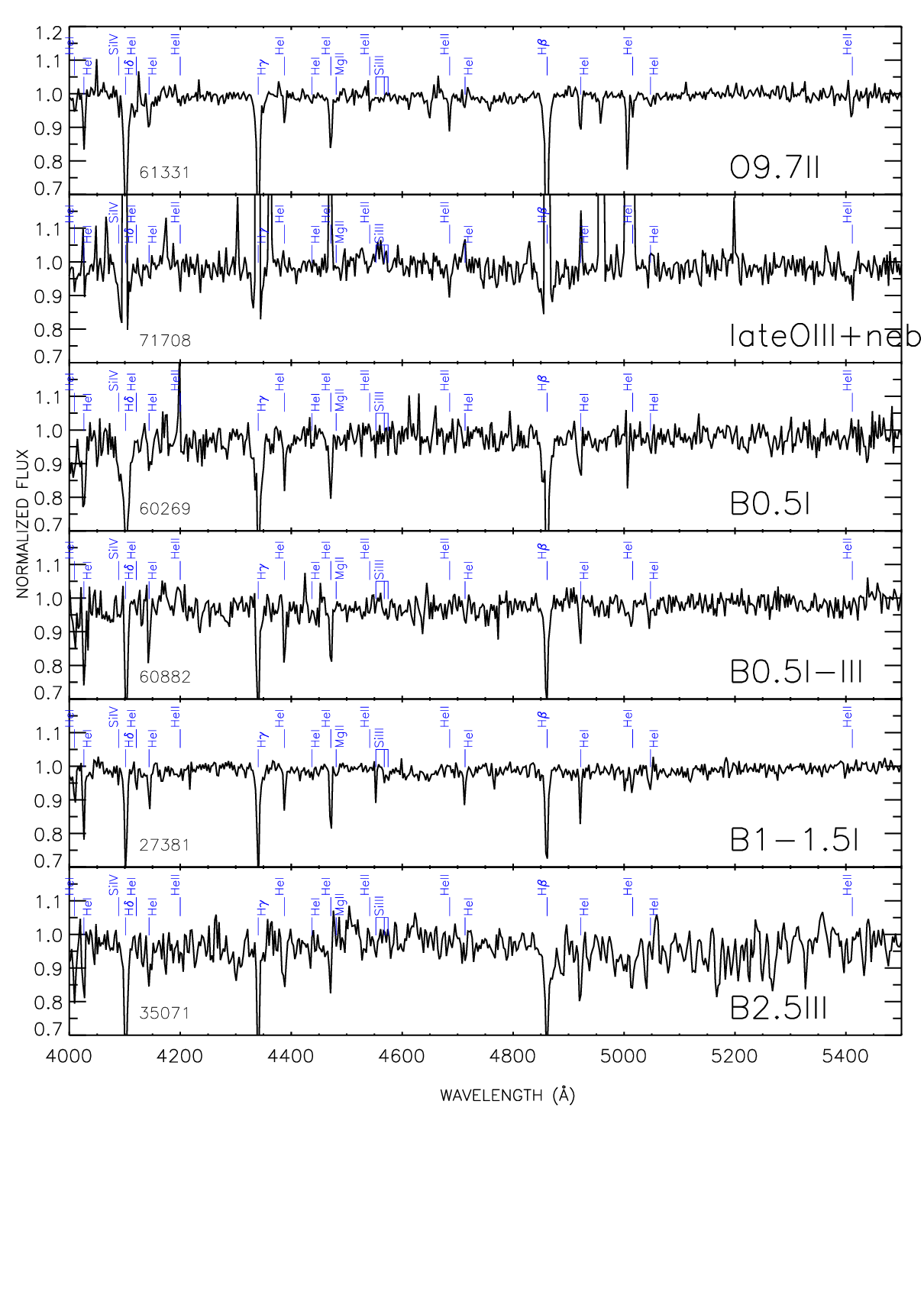}
   \caption{Fig.~\ref{F:all0}, continued. }
   \label{F:all1}%
\end{figure*}

The reduced stellar spectra are shown in Figs.~\ref{F:all0} to \ref{F:all1}.

%____________________________________________________________________________________________________

\section{Spectral Classification}        %___________________________________________________________
\label{S:SpT}

Spectral classification was made after \citet{WF90}'s and \citet{LDF92}'s
criteria for Milky Way stars bearing in mind metallicity effects,
as explained in \citet{Cal08} and summarized in \citet{C10}.
As far as we know, no spectral classification criteria
exists for metallicity smaller than the SMC.
%which we leave pending for future work when higher SNR spectra of
%a larger sample are available.
We avoided diagnostics involving simultaneously
metallic and non-metallic lines as much as possible.
They may be misleading due to the different
metallicity of IC1613 and the MW.
%Whenever possible no se usaron diagnosticos mixtos que usaran He lines
%and metal lines at the same time, since given the poorer 
%metallicity of this galaxy compared to the MW where the
%criteria are defined, this may be misleading.
Spectral subtypes for O-stars were derived comparing mainly the ratios of the lines HeII4541/HeI4471,
HeII4200/HeI+HeII4026, HeII4541/HeI4387 and HeII4200/HeI4144,
and luminosity class from the HeII4686 line and the SiIV4089/HeI4121 ratio.
The main diagnostics for B spectral subtypes were
SiIV4089/SiIII4552, MgII4481/HeI4471, SiIII4552/SiII4128, SiIII4552/MgII4481 and SiII4128/HeI4121,
and for luminosity class SiIV4089/HeI4121 and SiIII4552/HeI4387.
As the luminosity criteria often involve the undesired He-to-metal ratio,
we checked that the observed visual magnitude, corrected from extinction,
agrees with the calibrated for the assigned luminosity class.
The spectral classifications are provided in Table~\ref{T:assoc}.

The main sources of uncertainty are poor SNR, poorly removed or over-subtracted
nebular contamination, unremoved cosmic-rays (exposure times were long and only
2 exposures per target were taken) and the difficult normalization around \hdelta,
that affects all diagnostics involving SiIV4089 and HeI4121.
Additionally, the increased background levels
of the OBs executed in bright time hamper the
detection of the weakest spectral lines;
it is possible that weak HeII lines are not detected in actual late-O stars,
resulting in a bias towards B-types.
%Some OBs were executed in bright time and sky contamination 
%is critical for those targets.
%Not only the overal spectral SNR is poor,
%but also the increased background levels will hamper the
%detection of the weakest spectral lines.
%It is likely that weak HeII lines are not detected,
%resulting in a systematic error towards later spectral types in the spectral
%classification process.

We estimate that spectral types are uncertain by about 1-2 sub-types,
depending very much on the SNR of the spectrum and the nebular
contamination.
Nonetheless we are confident on the O and B classification:
HeII lines are clear in the O-types,
HeI and Si lines are clear in the
B-types,
and all the spectra show stellar wings in the Balmer lines.
%eventhough the specific diagnostic lines may be hindered by noise,
%HeII lines and the HeI and Si lines are clear in the
%O/B types respectively,
%and all the spectra show clear stellar wings in the Balmer lines.
As we have pointed out, the only risk in this sense
is that early-B stars may actually be late-O stars with
very weak HeII lines, undetected due to high background levels.
%Esto se hizo inicialmente con compare\_lambda.pro que esta en el nas4,
%y luego lo he repasado con ../ANALYSIS/SpTrev.pro con usando las velocidades radiales finales.

From the maximal discovery expectation of thirteen new O-type stars,
we have found eight O-stars and three early-B stars,           
and provided a similar spectral type for the formerly known early-O star 65426.
35071, whose colors where similar to the LBV-candidate V39,
turned out to be an additional B-type star.

The observations did not produce spectra apt for classification
%did collect enough photons 
for the three remaining stars.
68456, located close to IC1613's WO in the \qub~ diagram, is dominated by nebular lines
with no trace of WR's typical broad wind emissions.
44736 and 67063 exhibit Balmer lines, and seem to also show
HeI lines, but the spectra are too noisy for classification.
%No spectral types are provided for
%them in Table~\ref{T:assoc}

%In summary, we have produced sound spectral types for twelve new OB-stars in IC1613
%with only five slit pointings, 
%proving that the followed selection procedure works well.
%%This is quite remarkable considering that only five hours GTC observing time were devoted
%%to the program, that only one hour was devoted to each target,
%%that the observing conditions in some cases were poorer than demanded
%%by the proposal, and that most of the stars are fainter than V=19.

\subsection{Notes on individual targets:}
\label{SS:NiT}

\textbf{64066 (O3~III((f))):}   %\textbf{ob0003\_x1.3:}\\
The star displays strong HeII lines
whereas HeI lines can be hardly seen.
The spectrum also exhibits stellar Balmer wings with intense
nebular lines at the core.
We examined the subtracted sky
spectrum and
concluded that it seems unlikely that the
absence of HeI lines is caused by nebular contamination.
%The spectrum also exhibits stellar Balmer wings with intense
%nebular lines at the core, and weak NIII4634-40-42 emission.
Since NIII4634-40-42 is weak in emission,
the star is classified as O3III((f)).
There may be emission of NIV4058 in the spectrum
which would render the star an ((f*)),
and also of CIII4647-50-51;
however, the SNR is too poor to be conclusive.
64066 is very close to the O9~V star 63747, $\sim$0.5mag fainter,
but contamination is unlikely given the absence of
HeI lines in the spectrum of 64066.
%The star might be contaminated by the nearby star 63747,
%$\sim$0.5mag. fainter and O9~V type.

%Given the total absence of HeI lines (no hint of 
%absorption, nor emission), it seems unlikely that
%these lines are contaminated by the nebulosity.
%Finally, we subtracted sky spectrum is dominated
%by the local sky, caused by the bright moon.

\textbf{65426 (O6~III):}   %\textbf{ob0002a\_x2.4:}\\
This target is located inside nebulosity,
and the [OIII] lines indicate incomplete sky subtraction.
The star displays broad stellar Balmer wings with strong
nebular lines at the core. 
However, HeI lines do not seem contaminated by the nebula.
HeII4686 has an artificial P~Cygni-like profile
caused by a strong cosmic-ray in one of the exposures.
%, it could be the combination
%of exp.1 absorption plus strong emission (as wide as Balmer emission) in exp. 2.
We only consider the absorption part of the line.
The star was previously classified as O5-6~V by \citet{Fal07},
but we find a slightly later spectral type and higher luminosity class, O6~III.

\textbf{69476 (O6.5~III):}   %\textbf{ob0002a\_x2.2:}\\
The star is located inside GS1 \citep{MCW88}, one of the large bubbles of the galaxy,
and the sky subtraction is problematic.
The [OIII]5007 line indicates a slight sky undersubtraction
while the profile of the Balmer lines at the core hint that the nebular 
contribution has been oversubtracted.
HeI4387 may also experience nebular contamination,
but other HeI lines seem free from nebulosity.
The HeII lines are broader than HeI.
There may be emission of NIII4634 and CIII4647-50-51, which would render
the star an ((fc)) type, but SNR is too poor to be conclusive.

\textbf{36611 (O7~III-V):}   %\textbf{ob0005\_x2.3:}\\
This star was observed in bright time and almost at twilight,
which explains the comparatively poor SNR of its spectrum.
The [OIII] lines indicate severe nebular contamination,
and the spectrum displays strong
nebular lines at the core of the stellar Balmer lines.
%It displays broad stellar Balmer wings with strong
%nebular lines at the core. 
%Additional radial velocity. Normalization problems?
The HeII lines are broader than the HeI lines
and seem to have substructure.
%When extracting with \textit{apall} we found
%a very star 36894 1.5mag. fainter.  %%%%%% Esta no deberia contaminar porque esta a 3segundos de arco.
The GHV09 catalog has two very close sources
about 1\arcsec~ to the SW, which are fainter by 0.8 and 2.3 magnitudes.
We mark 36611 as possibly multiple.

The P~Cygni-like shape of HeI4471 is caused by a 
not totally removed cosmic ray
in one of the exposures.
HeII4686 is a bit stronger than HeII4541 indicating class-V, but
the SiIV4089/HeI4121 ratio suggests class-III.
The star's absolute magnitude, if single,
is also consistent with an intermediate luminosity class.

\textbf{75661 (O8~III):}   %\textbf{ob0002a\_x1.1:}\\
The normalization of this star is very problematic
and hampers spectral classification.
The [OIII]5007 nebular emission indicates that the nebular extraction
was not complete.
The spectral diagnostics disagree: the HeII4686/HeII4541 ratio clearly indicates
that the star has spectral type earlier than O8, but
HeII4541/HeI4471, HeII4200/HeI4144 and HeII4200/HeI4026 point towards later spectral types.
There seems to be MgII4481 in the spectrum.
The weak HeII4686 could be due to a relatively strong wind filling the line, which would agree 
with the relatively weak H$_\beta$ line.
The assigned type is a compromise of these pieces of evidence
but the possibility that the star is a binary,
specially considering the difficulties in determining its radial velocity, is high.
The assigned radial velocity -250\kms matches well HeII4541, HeII4686, \hbeta, HeI4922
and HeI4387, but \hgamma, HeI4471 and HeII4200 indicate vrad=-350\kms.
%%There may also be a problem with the wavelength calibration for this star.
%%\textbf{por que? y por que es tan ruidosa?
%%es la secundaria del run!! Esta en el borde del chip?} --> porque tiene magnitud 19.8 !
%%Normalization problems (\textbf{revisar esto a la vista del plot final}).
%%\textbf{Medir vrad}, \hbeta~  desdoblada? (bueno, esto es por la neb.cont. a la vista del OIII)
The luminosity class was derived extrapolating criteria from the earlier O6 types;
the star's absolute magnitude is close to the value calibrated for O8 supergiants
by \citet{winter}.

\textbf{67684 (O8.5~I):}   %\textbf{ob0004\_x2.5:}\\
The [OIII] lines in absorption indicate that the sky was oversubtracted,
which explains the strong artificial absorption at the core of the Balmer series.
The spectral SNR is poor, as the star was observed at large
airmass and bright time conditions.
The normalization of the spectrum was problematic and hinders spectral classification;
He lines are apparently found in broad bands
with overlapped narrow lines. HeII4686 looks asymmetric and 
with several components.
%All this indicate the star may be binary.
The spectral SNR prevents us from deciding whether this is an artifact 
from normalization or the star is multiple or
actually a compact cluster. Nonetheless HeII4686 and HeII4200 are strong.
We mark it as possibly composite spectrum.

\begin{figure*}
\centering
   \includegraphics[width=\textwidth]{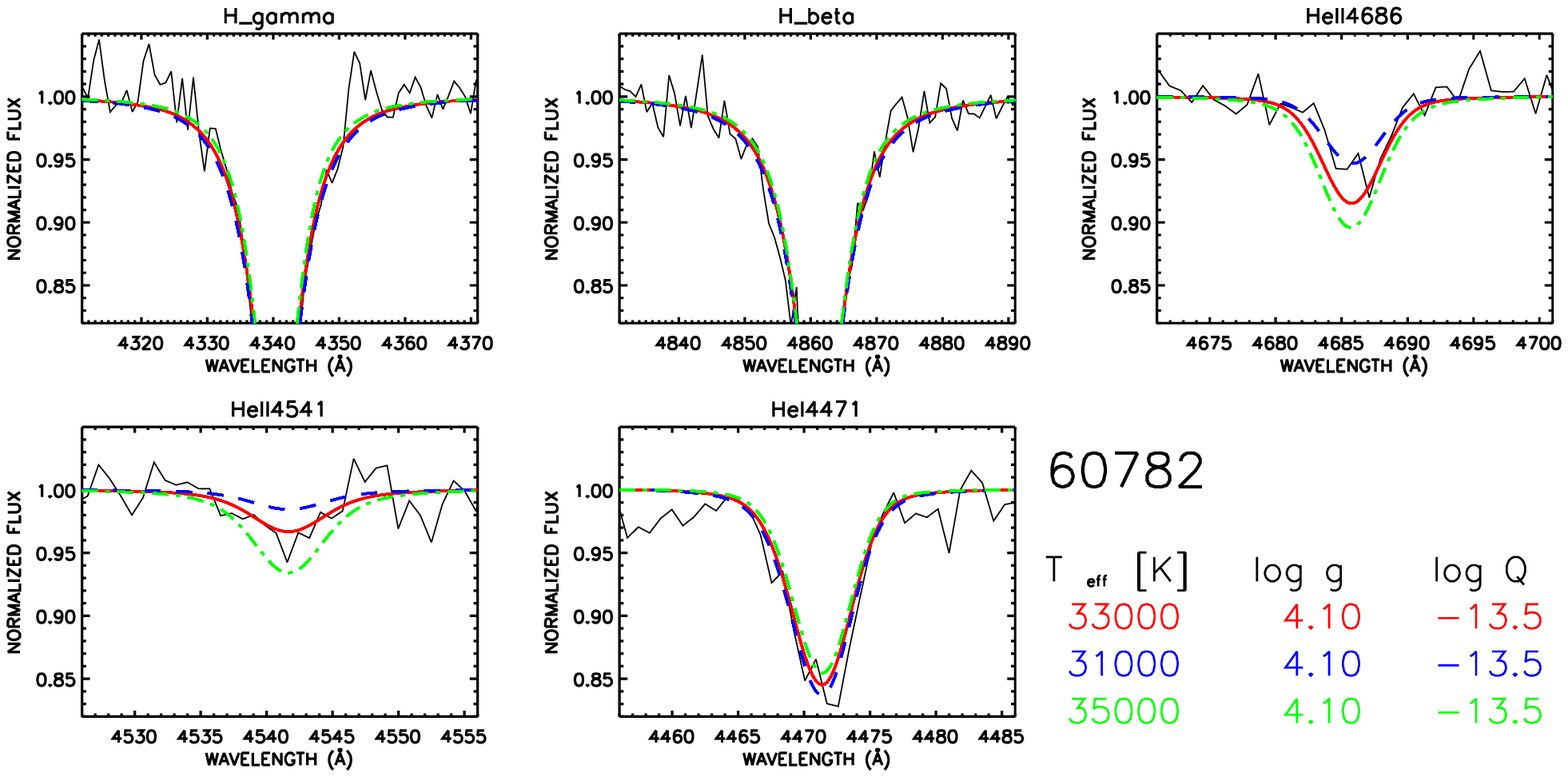}
   \caption{Best fit FASTWIND model (red, solid line) to sample star 60782 (black).
The chart shows the principal diagnostic spectral lines: \hgamma~ and \hbeta~ (gravity),
HeII4541 and HeI4471 (\Teff), and HeII4686 (\logq).
To illustrate that \Teff~  is well constrained within the error bars,
two additional models (non-solid lines) with \Teff~ varying $\pm$ 2000K are shown.
The fit of these two models to the HeII4541 diagnostic line is poor,
defining the error bars for temperature.
}
   \label{F:fit}%
\end{figure*}

\textbf{60782 (O9.5~III):}   %\textbf{ob0001a\_x1.4:}\\
The star is located in a region of nebulosity with intense
emission in the Balmer lines.
The [OIII]5007 line indicates a slight sky oversubtraction.
The spectrum may be contaminated by 61331,
which is very close and 0.5 mag brigther
but 60782's wider Balmer lines and spectral morphology in general suggest otherwise.
There is also a nearby star with very similar V-magnitude and Q-color (60794, V=19.64, Q=-0.99),
but it should be left out of the slit at the chosen position angle of the observations.
During spectral extraction with \textit{apall} additional fainter nearby stars
where detected.
%This star is marked as 'multiple? blend?' in my reduction notes.

SiIV4089 is  blended with \hdelta~ difficulting luminosity classification.
%HeII4541 indicates type O8, but all other diagnostics 
%point to O9.5.

\textbf{61331 (O9.7~II):}   %\textbf{ob0001a\_x1.3:}\\
The star is located in a region of nebulosity with intense
emission in the Balmer lines.
The [OIII]5007 nebular line indicates sky oversubtraction,
which explains the artificial absorption at the core of the Balmer lines.
The apparent emission of HeII4541 is due to a cosmic-ray in
one of the exposures.
The star is very close to 60782 but contamination seems unlikely as
the later is 0.5mag fainter.
The lines of hydrogen, HeI and HeII yield different radial velocities.
The apparent mismatch between HeI and HeII is probably due to a poorly removed cosmic ray
in the wing of HeI4471, and the one at HeII4541.
The adopted radial velocity was calculated with HeI lines.
The HeII4200 line is very weak.
The luminosity indicator SiIV4089/HeI4121 suggests class 
between I and III, hence the assigned class-II. The star's absolute magnitude,
calculated from the apparent magnitude corrected by extinction,
is also consistent with an intermediate class.
%Note, however, that \citet{Sota11}'s HeII4686/HeI4713 criterion
%would suggest class-V.

\textbf{71708 (late-O~III + neb.):}   %\textbf{ob0002a\_x2.1:}\\
%Intended as prio-C in slit\_b
Spectral quality is too poor to derive radial velocity;
IC1613's systemic velocity -234 $km \, s^{-1}$ \citep{Lal93}
is adopted.
The spectrum displays broad stellar Balmer wings with strong
nebular lines at the core.
HeII lines, specially HeII4541 and HeII4686 are clearly seen
and indicate an O-type. No spectral type can be assigned, however,
since the spectrum displays no stellar HeI lines, very likely due
to nebular contamination. SiIII4552 is not seen in the spectrum.
HeII4686 is strong in absorption, and stronger than HeII4541;
according to the luminosity criteria, this corresponds to luminosity
classes III and V for spectral types O8-O9.
Its absolute magnitude agrees with the $\rm M_V$~ calibrated for 
a late-O giant \citep{winter}.

\textbf{60269 (B0.5~I):}   %\textbf{ob0001a\_x1.5:}\\
The [OIII]5007 line indicates sky oversubtraction,
which explains the asymmetry and strong absorption at the core of the Balmer lines.
There is a velocity shift between the nebular lines
(at IC1613's systemic velocity, measured from the sky spectrum) and the 
Balmer absorption from the final spectrum.
HeI and metal lines experience a smaller shift.

This star was not a primary target of the observing run;
it is very faint (V=20.5) and consequently its spectrum is noisy. 
All B-subtype diagnostics point towards an early-B type,
but the absence of HeI4121 hints B8-9 types.
However, it is possible that HeI4121 is contaminated
by nebular emission and this diagnostic was discarded.
In absence of HeII4686 and HeI4121 lines,
the luminosity class was assigned only from the SiIII4552/HeI4387 ratio,
resulting in class-I. 
However the Balmer lines are too broad for a supergiant: the star
may be a late-O type star with undetected HeII lines due to
high background level compared to the stellar spectrum signal.
Its absolute magnitude ($\rm M_V \lesssim -4$) is consistent with the star being a
late-O or an early-B dwarf star, but not with a supergiant ($\rm M_V = -6$).
The possibility that its spectrum is contaminated by
the very nearby star 60782 (O9.5III) which is 1mag brigther cannot be discarded.

\textbf{60882 (B0.5~I-III):}   %\textbf{ob0005\_x1.1:}\\
The [OIII] lines indicate a correct background extraction,
although the profiles of the Balmer series core indicate a slight oversubtraction.
This star was observed in bright time and almost at twilight,
which explains the comparatively poor SNR of its spectrum.
No HeII lines can be seen except perhaps HeII4541;
the high background level of the exposures
could be hiding the presence of weak HeII lines but the SiIII triplet
is strong and suggests a B-type.
The SiIV4089/HeI4121 ratio indicates class-I
but SiIII4552/HeI4387 suggest class-III.

% \textbf{27381:}   %\textbf{ob0001a\_x2.1:}\\
% no tiene comentarios

\textbf{35071 (B2.5~III):}   %\textbf{ob0003\_x2.3:}\\
The star seems double at OSIRIS acquisition image,
which has enhanced resolution compared to GHV09's INT-WFC catalog.
The criteria based on MgII4481, HeI4471 and SiIII4552
point towards types B2.5-3.
We note, however, that the presence of SiII4128,
expected at this spectral sub-type at higher metallicities, is not clear.

\section{Spectroscopic analysis}                       %___________________________________________________________
\label{S:FW}

The resolution and SNR of the dataset enabled a rough 
determination of the stellar parameters of the sample O-stars.
We used synthetic spectra calculated with
FASTWIND version 10.1 \citep{FW05}
that takes into account line blanketing, non-LTE
effects and radiation driven winds,
to fit mainly \hgamma, \hbeta, HeII4686, HeII4541 and HeI4471,
and secondarily HeII4200, HeII5411, HeI4387 and HeI4922.
%to fit optical spectral lines of hydrogen and helium:
%\hgamma, \hbeta,                       %\hgamma~ (main), Hbeta,                          
%HeII4686,                              %HeII4686,                                      
%HeI4471, HeI4387, HeI4922              %HeI4471 (main), HeI4387, HeI4922               
%and HeII4541, HeII4200 and HeII5411.   %and HeII4541 (main), HeII4200 and HeII5411.     

The synthetic spectra were taken from a 
vast grid of FASTWIND models with metallicity 0.13\Zsun,
computed using the CONDOR facilities at the IAC \citep{SSal11}.
0.13 \Zsun~ is the equivalent metallicity 
to the oxygen abundance derived by \citet{Fal07} for B-supergiants in IC1613.
The grid covers from 28000 to 55000K,
gravity ranging $\rm \log \, g=$2.6-4.3 (depending on temperature)
and wind-strength Q-parameter $\rm \log \, Q=  \dot M / (v_{\infty} \, R_{\ast})^{1.5}$  from $-$15.0 to $-$11.7.
The exponent of the wind velocity law $\beta$,
the helium abundance and the microturbulence also vary in the grid,
but were kept constant to typical values for this analysis:
$\beta$=0.8, $\epsilon _{He}$=0.09 and $\xi$=10\kms.

Starting from effective temperature and gravity (\logg) values typical for the star's spectral
type, 
we varied \Teff~ in 1000K steps to fit simultaneously 
HeII and HeI lines.
We then fixed \Teff, and varied \logg in 0.1dex steps to fit the wings of the Balmer lines
(the core was discarded since nebular contamination was severe,
or unquantifiable, in many cases).
The best-fitting gravity value usually requires the recalculation of \Teff.
Therefore, we iterated this process until all considered spectral lines
except for HeII4686 were fitted.
The observations do not include \halpha, but HeII4686 also provides information on the wind.
We fitted this line by changing the wind strength \logq parameter.
If the wind is not negligible it alters other lines under analysis, 
and the whole process must be iterated again to adjust temperature and gravity.
Only when  HeII4686 displays a clearly non-photospheric profile (e.g. asymmetric), 
we allowed $\beta$ to vary.

\begin{figure*}
\centering
   \includegraphics[width=\textwidth]{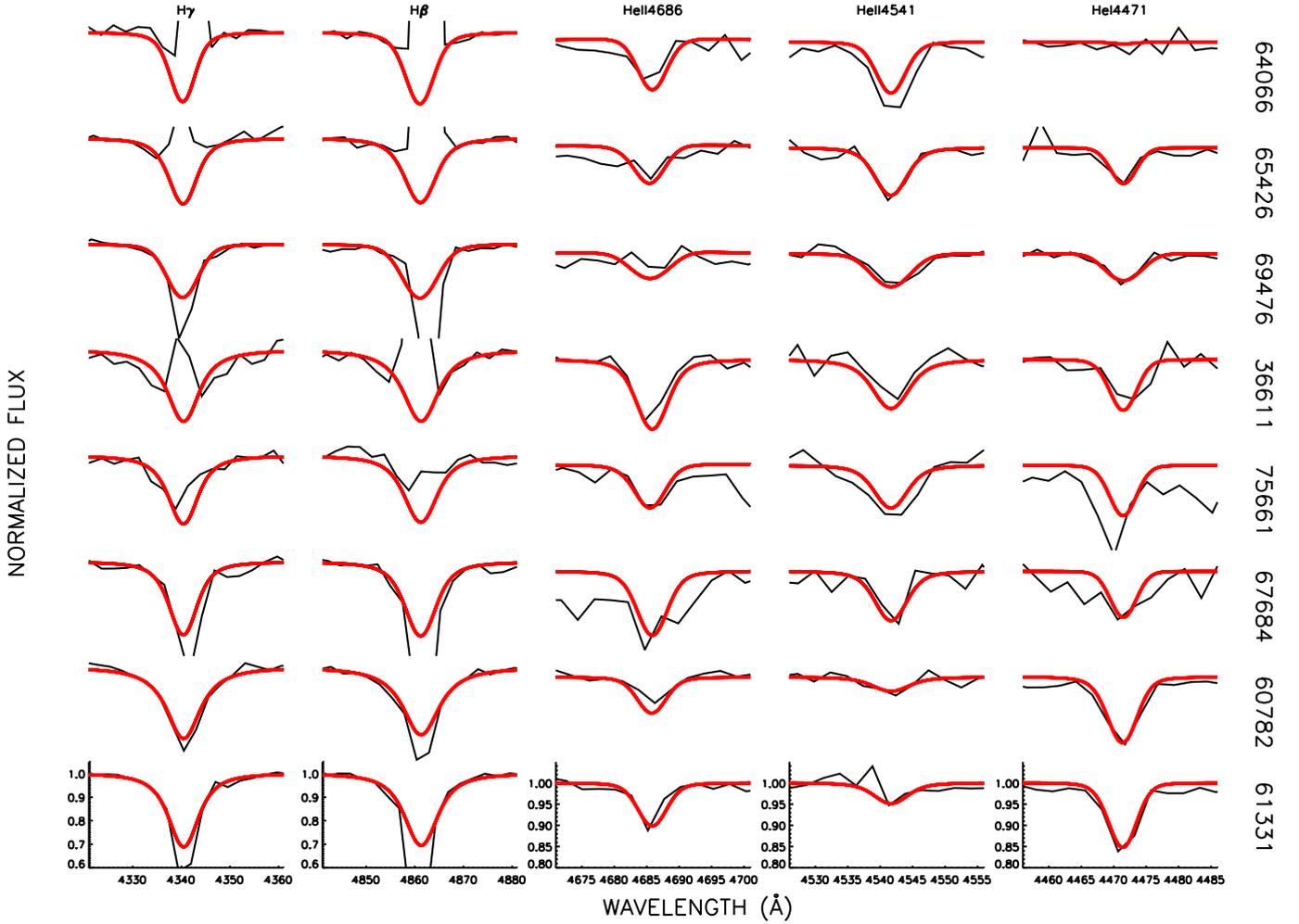}
   \caption{Best fit FASTWIND model (red) to the sample stars (black).
The chart shows the principal diagnostic spectral lines, as in Fig.~\ref{F:fit}.
Stellar identification numbers are provided on the right-hand side of the plot.
Axes are provided only for the bottom target for clarity sake;
the same axes are used for the remaining stars.
}
   \label{F:fitall}%
\end{figure*}

%Never use the core of line for analysis. Both smooth
%and my rebin routine make the lines shallower.
%Looked for other rebin routines with flux conservation
%but did not find any, or did not work

%We kept beta (0.8), microturbulence (0.10) and He-abundance (0.09) constant,
%since the spectral quality does not enable to contrain this parameter.
%Hay que justificar estos valores. Para beta puedo coger las RFs de  la propuesta ESO.
%para micro, el paper de Charo. Y para He-abu, sera laestandar.
%Only when  HeII4686 is strange (e.g. asymmetric), we varied the value of beta.

At the spectral resolution of the dataset, rotational
and macroturbulent velocities could not be derived
unless they are as high as $\sim$250\kms.
We also measured radial velocities for the targets.
We noted some problems with the wavelength calibration, specially in the red part of the spectrum,
that affected the HeII5411 line. 
This cannot be explained by the shifts detected between consecutive exposures,
but rather to the reported calibration problems in the red part of some spectra (Sect.~\ref{S:obs}).

%\textbf{Fig.~\ref{F:fit} illustrates the goodness of the fits,
%leading to effective temperature estimates typically within $\pm$2000 K}
%Given the spectral quality of our data,
%we can narrow down the effective temperatures estimates to within $\pm$2000 K

The spectral quality of the dataset enabled the determination of
effective temperatures within $\pm$2000-3000 K,
as shown in Fig.~\ref{F:fit}.
Typical errors bars for \logg are $\pm$0.2.
We do not quote errors for \logq as the method provides only a rough estimate of
the wind influence on other stellar parameters, and the derived values must be considered orientative.
%The spectral resolution of the dataset prevents us from deriving rotational
%velocities and macroturbulent velocities, unles they are of the order of
%300km/s.
%Only if strange broadening of one set of lines, I play with vmacro.
%We also derived radial velocities for the targets, but could not use
%the same line for all the stars because of cosmic rays, nebular contamination, noise,...
%We noted a problem in calibration for red wavelengths
%that rendered the 5411 line unusable.

The spectral fit to the sample O-stars is shown in Fig.~\ref{F:fitall}.
Results are presented in table \ref{T:fits}.
\begin{table}
\caption{Parameters derived for the sample stars }           
\label{T:fits}      
\centering        
\begin{tabular}{l l l l l l}     
\hline\hline  

     ID &         SpT &          vrad &                           \Teff &\logg & \logq \\
    GHV09&           & $km \, s^{-1}$ &                               K &      &       \\
\hline                                
 64066 &  O3III((f)) & -240 & $\geq$ 49000                    & 3.8  &-12.7 \\
 65426 &       O6III & -270 &        40000 $ \rm \pm  2000  $ & 3.8  &-12.5 \\
 69476 &     O6.5III & -350 &        37000 $ \rm \pm  2000  $ & 3.4  &-12.5 \\
 36611 &     O7III-V & -140 &        40500 $ \rm \pm  3000  $ & 4.3  &-15.0 \\
 75661 &       O8III & -250 &        37500 $ \rm \pm  3000  $ & 3.9  &-12.7 \\
 67684 &       O8.5I & -260 &        38500 $ \rm \pm  3000  $ & 3.8  &-15.0 \\
 60782 &     O9.5III & -190 &        33000 $ \rm \pm  2000  $ & 4.1  &-13.5 \\
 61331 &      O9.7II & -180 &        33000 $ \rm \pm  2000  $ & 3.8  &-15.0 \\

\hline
\end{tabular}
\end{table}

\paragraph{Comments on individual targets}
~\\

\textbf{64066 (O3III((f))):}
This star displays only absorption lines of HeII,
with no trace of HeI.
We were able to derive only a lower limit for effective temperature.
Consequently gravity is also poorly constrained.

%\textbf{65426 (O6III)}  
%suffers from strong nebular contamination in Balmer series that makes 
%the derived \logg uncertain.
%The wavelenth calibration is unreliable at 5411\AA,
%and maybe also from 4200\AA~ bluewards.
%The stellar HeII5411 line is stronger than HeII4541, and is not
%well fitted by the models.

\textbf{69476 (O6.5III):}
\hbeta~  could not be used for analysis due to nebular contamination.
The helium lines are very broad: we adopted vsini=250\kms.
The HeII4541 and HeII4686 lines indicate vrad$\sim$-280\kms,
but HeI and Balmer transitions yield $\sim$-350\kms.
69476 is inside the bubble GS1, and the different radial velocities
could be artificial and caused by undetected nebular contamination 
of gas masses moving at different radial velocity than the star. 
However, we examined the sky spectrum
and there seems to be no nebular emission in HeI or HeII lines.
Another possible explanation is that the star is a binary.
We adopted the radial correction calculated with HeI lines.

%\textbf{36611 (O7III-V):}
%Its derived high gravity fits the He lines
%and \hbeta, but \hgamma~ is still broader than the models. 
%This could be a normalization problem, specially difficult in the \hgamma~ region,
%or that the spectrum is composite.
%The radial velocities derived from HeI and HeII lines disagree.
%As we pointed out in Sect.~\ref{SS:NiT},
%this target may be blended with a nearby star.
%The discrepant radial velocities also hint that the spectrum is composite
%and/or the star is a binary.

%\textbf{75661 (O8III):}
%The radial velocity derived from HeI and HeII lines is slightly different.
%The spectral SNR, the uncertain spectra and the nebular contamination
%render \logg (and indirectly \teff) hard to contrain.
%The stellar spectrum displays stronger absorption
%lines of both HeI4471 and HeII4541 than the model, indicating enhanced helium abundance.
%The HeI4471 line is found at slightly different lambda
%than expected but other HeI lines are found where expected, 
%indicating perhaps a problem with the wavelength calibration.

%\textbf{67684 (O8.5I):}
%As explained in Sect.~\ref{SS:NiT} the spectral quality is poor,
%the continuum level is uncertain, and the spectrum may be composite.
%This also applies to the HEII4686 line,
%which makes the Q-parameter poorly contrained.

\textbf{60782 (O9.5III):}
The best fit to HeII4686 would actually require \logq=-13.0, 
but Balmer lines could not be fitted with this strong wind
even if \logg~ is increased.
Contamination by nearby objects (see Sect.~\ref{SS:NiT}) could be a possible explanation.

\textbf{61331 (O9.7II):}
%(T320g360He09Q150b08 tb es bueno, pero T325g370He17Q150b08 better)
HeII4686 is blue-shifted and narrow, and is not reproduced
by any combination of $\beta$ and \logq.
%The core of the HeI lines could not be fitted;
%by comparing the model and the observed spectral lines,
%we conclude that the fit would not improve if increasing the He-abundance.
%A possible explanation for this mismatch is the oversubtaction
%of nebular He lines.
%The star may be Helium-overabundant:
%the line of HeI4471 cannot be fitted even decreasing down to 28000K,
%in contrast with other helium lines.

%\textbf{B-stars: 60269 (B0.5I), 60882 (B0.5I-III), 27381 (B1-1.5I),
%35071 (B2.5III).}
%They display no HeII lines, and we cannot use the He ionization
%equilibrium to constrain temperature.
%Only upper limits for temperature, given by the ocurrence of HeII lines,
%can be set.
%Consequently, gravity is also poorly derived.

\subsection{The very low metallicity temperature scale}

\begin{figure}
\centering
\includegraphics[width=0.45\textwidth]{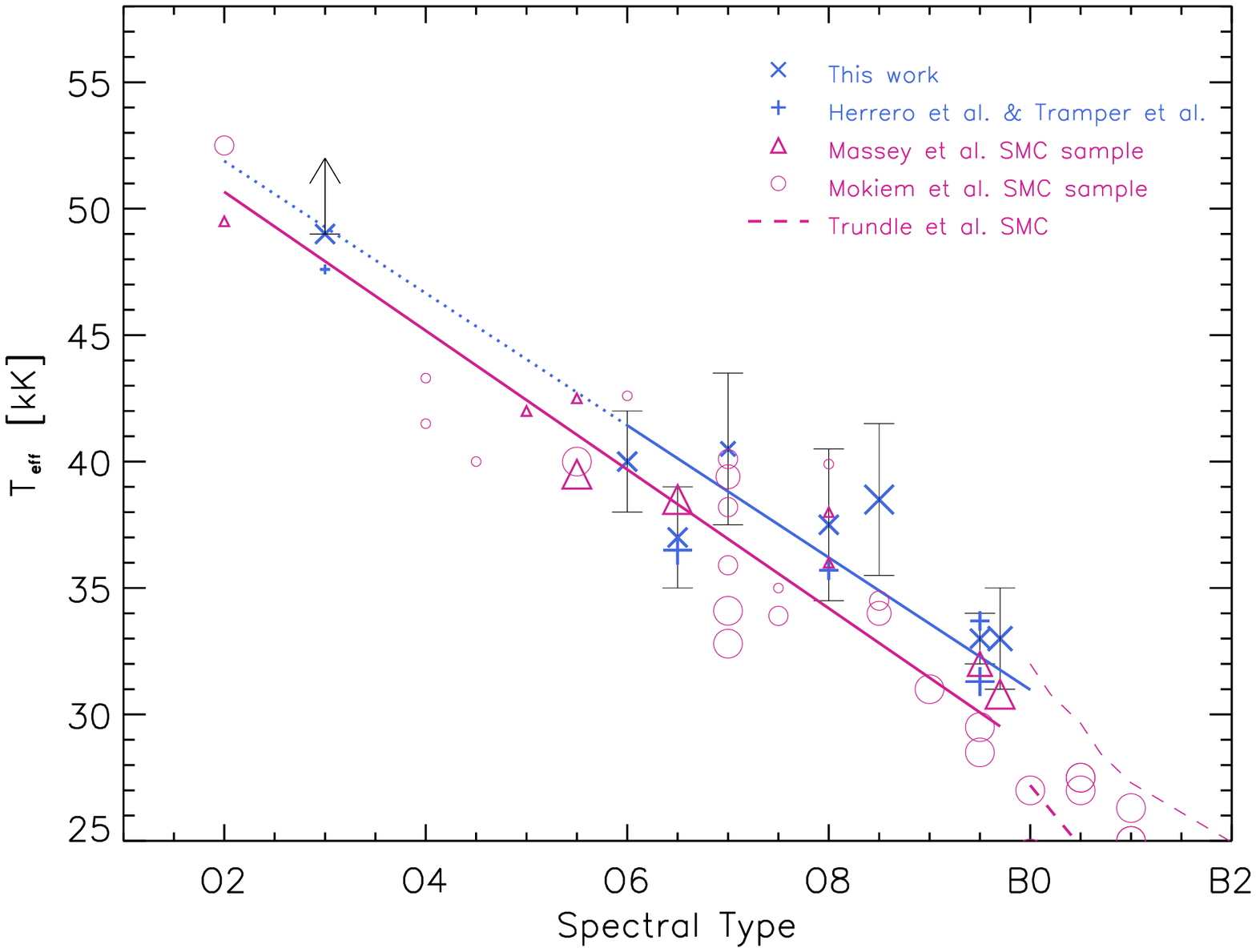}
   \caption{Very low metallicity temperature scale for O- and early-B stars, 
including this paper's results and the stars analysed by
\citet{Tal11} and \citet{Hal12} in IC1613.
Different symbol size indicates different luminosity class (smallest symbols = class~V,
largest symbols = class~I).
For comparison, we have included derived temperatures for OB-stars 
in the SMC from \citet{Mal07} and \citet{Mal09} (violet),
and \citet{TDH07}'s calibration for SMC B supergiants
(dashed-thick) and dwarfs (dashed-thin).
%The solid line represent linear regressions to the joint 
%sample of \citet{Mal07} and \citet{Mal09} SMC giant and superg
Solid lines represent linear regressions to the joint 
sample of O-giants and supergiants in the SMC (violet)
and IC1613 (blue).
The high end of the temperature scale in IC\,1613 is not well constrained, 
hence marked with a dotted line.
%We also include as reference the calibration of \citet{Mal05b} for Milky Way stars.
           }
      \label{F:Teff}
\end{figure}

We present the first sub-SMC temperature scale in Fig.~\ref{F:Teff}.
It includes all O-type stars in IC1613 whose effective
temperature has been derived from quantitative spectroscopic analysis:
the 8 stars analyzed in this paper,
plus 5 stars from \citet{Tal11} and \citet{Hal12}.
For comparison, we have added temperature determinations of SMC OB-stars
from \citet{Mal07} and \citet{Mal09}, and the temperature
scale for SMC B-supergiants derived by \citet{TDH07}.
%As we pointed out in [GHC10], the SMC data display the expected increasing
%trend of effective temperature with earlier spectral types, but the
%trend is less steep for the O-types than for B-types.
%Although we have not derived parameters for B--types, our results for O stars in IC\,1613
%indicate a similar (scaled) behaviour than stars in the SMC. 
At the same spectral type, the temperatures we derive for IC1613
stars are similar to SMC stars, eventhough slightly hotter.

%The temperatures we derive agree with those found for the same
%spectral types for SMC stars eventhough slightly hotter,
%as expected given IC1613's poorer metal content.
%\textbf{
%%It may be argued whether the difference is significant
%%given the error bars of all data points.
%%and whether it accounts for the expected gap
%%These points can only be addressed 
%%after the quantitative analysis of enhanced
%%resolution spectra (R$\sim$5000) produces
%%the stellar metallicity and a more accurate temperature determination.
%Whether the difference is significant
%can only be addressed with the quantitative analysis of enhanced
%resolution spectra (R$\sim$5000) to produce
%the stellar metallicity and a more accurate temperature determination.
%}
%It may be questioned whether the difference in temperature matches the
%theoretical expectation considering the difference in metallicity.
%This point, however, cannot be discussed until an individual 
%detailed metallicity analysis of the sample
%stars of this paper is performed.

We calculated a least squares linear regression
to the temperatures of the total sample of analyzed giants and supergiants in IC1613.
We proceeded similarly with SMC stars  (\citet{Mal07} and \citet{Mal09}).
The high end of IC1613's temperature scale (marked with a dotted line) was not
well constrained,
since we could only derive a lower temperature limit for the giant O3 of the sample.
The other O3 star \citep[from][]{Tal11} is a dwarf and was not considered in the fit.
As a test, we discarded the three stars marked as binary/multiple in Table~\ref{T:assoc} and obtained
a very similar temperature scale: uncertain binarity may introduce some scatter in the
derived relation but does not change it globally.

The derived scale is $\sim$1000~K hotter for IC1613 than for the SMC, as expected given IC1613's poorer metal content.
It may be argued whether the difference is significant
given the error bars of all points considered: $\sim$2500~K in this work
\textit{vs} typically $\sim$1000~K, but reaching up to $\sim$2000~K for some targets, in \citet{Mal07}.
It is in our programmed future work to produce a sound temperature scale for IC1613
from enhanced resolution and SNR spectra of a larger sample of stars in this galaxy.

%We have also calculated a least squares linear regression
%to the total sample of giants and supergiants in IC1613
%and the SMC (\citet{Mal07} and \citet{Mal09}).
%We note here that the high end of the temperature scale (marked with a dotted line) is not
%well constrained, 
%since we could only derive a lower limit of temperature for the giant O3 of our sample,
%and the other O3 \citep[from][]{Tal11} is a dwarf star not considered in the fit.
%If the three stars marked as binary/multiple in Table~\ref{T:assoc} were discarded, we would obtain
%a very similar temperature scale: uncertain binarity may introduce some scatter in the
%derived relation but does not change it globally.
%The derived scale is hotter for IC1613 than for the SMC, as expected.
%%If SMC and IC1613 temperatures are compared with \citet{Mal05b}'s calibration
%%for Milky Way early-type stars, the difference is substancially larger.

%\subsection{Singlet versus triplet lines}
%The problem (Najarro et al. 2006).
%There is a FeIV line close to HeI584\AA (vacuum wavelengths),
%that absorbs photons an ultimately drain the population of HeI atoms
%in the $\rm 1s \; 2p^1 \; P^0$. This affects other spectral
%lines to/from that level, such as 4389\AA~ ($\rm 1s \; 2p^1 \; P^0$~--$\rm 1s \; 5p^1 \; D$)
%or 4923 ($\rm 1s \; 2p^1 \; P^0$~--$\rm 1s \; 4d^1 \; D$).
%The authors discourage the use of singlet lines involving level $\rm 1s \; 2p^1 \; P^0$~ for analysis:
%4471 is preferred over 4387 and 4922.

%The effect decreases if the iron abundance decreases. It should not be
%seen in low metallicity environments, and Hillier \& Lanz (2001)
%did not find it when analysing SMC stars.
%The effect is more accute for increased microturbulence,
%as there is more overlap between FeIV and HeI587, hence more drain of the
%$\rm 1s \; 2p^1 \; P^0$ level. The singlet lines will decrease.

%Sp. types affected: O-dwarfs, mid-O supergiants (it doesn't say anything
%of early-O supergs). It doesn't affect B-stars, late-O supergs.

%Villamariz \& Herrero 2000: no derivan microturb.
%Encuentran que la diferencia entre meter microturb. o no afecta a los
%resultados pero dentro de las barras de error.
%Cuanto mayor es logg, menos se nota el variar la microturb.
%Lineas afectadas: HeI tanto en core como en alas,
%HeII 4686 solo core. Resto HeII y Balmer, nada o casi nada.

%____________________________________________________________________________________________________

\section{Summary and future work}        %___________________________________________________________
\label{S:con}

Very low metallicity massive stars are growing more and more
important to our understanding of the high-redshift very metal-poor Universe.
Current models for the first generations of stars rely on knowledge 
of present-day massive stars, with radiation driven winds being a main pillar.
While the paradigm has been thouroughly tested in the Milky Way and the Magellanic
Clouds, recent findings indicate that a different wind driving mechanism
may be at work at poorer metallicities.
The evolutionary models for population-III massive stars,
and the estimates of their ionizing, mechanical and chemical feedback
would have to be recomputed.
Only the quantitative spectroscopic analysis of a large
sample of very metal poor massive stars 
can shed new light on these topics.

This work uses the 10m telescope GTC as a Local Group explorer,
to unveil new OB-type stars in metal poor environments beyond the SMC.
From low-resolution long slit spectroscopy, 8 O-stars plus
4 early-B stars have been discovered.
The total list of known O-type stars in IC1613 
has been increased by a factor of 2.
A photometric-based selection method optimized towards O stars
has been presented, which will save observing time
to similar surveys in other galaxies.
The extension of the sample of very metal poor O stars 
is a fundamental first step to test the theory of radiation driven winds at sub-SMC metallicity.

The GTC-OSIRIS spectra were analyzed with FASTWIND models
to derive effective temperatures and gravities.
From the results we have produced the first 
temperature scale for O-type stars beyond the SMC.
The calibration yields higher temperatures for IC1613 stars,
as expected given their comparatively poorer metal content,
but must be used with caution given the large error bars of our results.

Follow-up spectroscopy of increased resolution 
and enhanced SNR is planned for a subset of this paper targets.
Its quantitative analysis will provide more accurate stellar
and wind parameters, and constraints to the wind theory.
The sample will be as extensive as possible in its
spectral type and luminosity class coverage
so that new clues on the strong wind problems are found.
Two interesting by-products will be produced:
the first atlas of very metal-poor massive stars and 
a sound sub-SMC effective temperature calibration.

\begin{acknowledgements}
This work has been funded by Spanish MICINN
under Consolider-Ingenio 2010, programme grant
CSD2006-00070, 
(http://www.iac.es/consolider-ingenio-gtc/),
and grant AYA2010-21697-C05-04,
and by the Gobierno de Canarias (PID2010119).
We would like to thank A. Cabrera-Lavers for
fruitful interaction during data processing.
This research has made use of Aladin,
and also data from the INT telescope 
(operated by the Isaac Newton Group in the Spanish Observatorio del Roque de los Muchachos).
We would like to thank our anonymous referee
whose comments helped us to improve this paper.
\end{acknowledgements}

%%%%%%%%%%%%%%%% BIBLIOGRAPHY %%%%%%%%%%%%%%%

%%%%%%%%%%%%%%%%%%%%%%%%%%%%%%%%%%%%%%%%%%%%%

\Online 
\begin{appendix}

%%%%%%%%%%%%%%%%%%%%%%% F_CHARTS %%%%%%%%%%%%%%%%%%%%%%%%%%%%%%%%

\section{Finding charts for the sample stars}
\label{S:f-charts}

Finding charts for the newly identified
OB stars are provided in this section. 
The images were taken with the Wide Field Camera (WFC)
at the 2.5m Isaac Newton Telescope (INT); for more details see GHV09.
All the identification charts show a field of $\rm 2 \arcmin \times 2 \arcmin$,
observed with broad Harris V-band filter.

\begin{figure}
\centering
\includegraphics[width=0.50\textwidth]{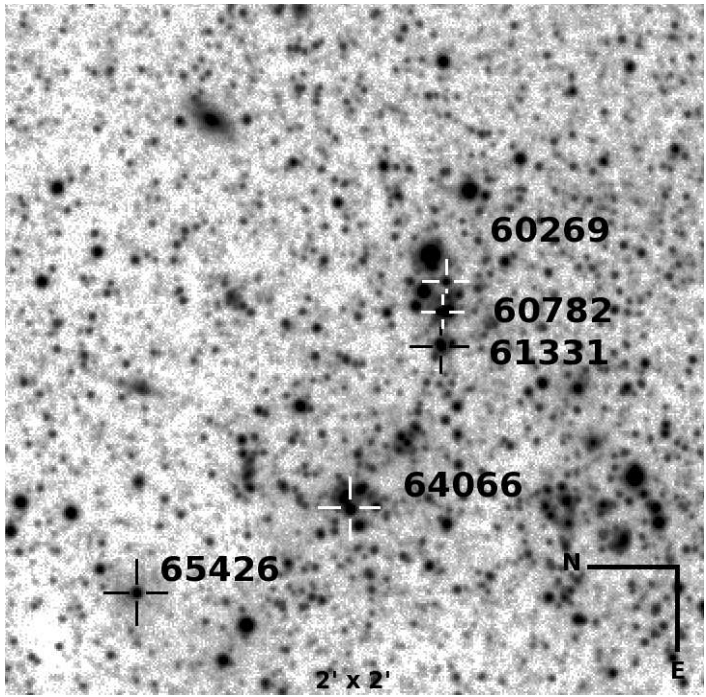}
   \caption{Stars 60269 (B0.5~I), 60782 (O9.5~III), 61331 (O9.7~II), 64066 (O3~III((f)))
and 65426 (O6~III).
           }
      \label{F:fc1}
\end{figure}

\begin{figure}
\centering
\includegraphics[width=0.50\textwidth]{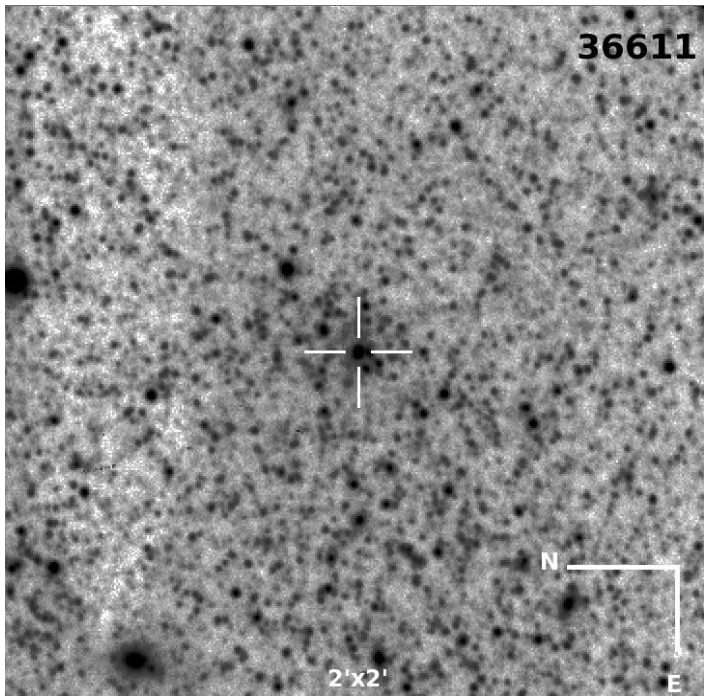}
   \caption{Star 36611 (O7~III-V).
           }
      \label{F:fc3}
\end{figure}

\begin{figure}
\centering
\includegraphics[width=0.50\textwidth]{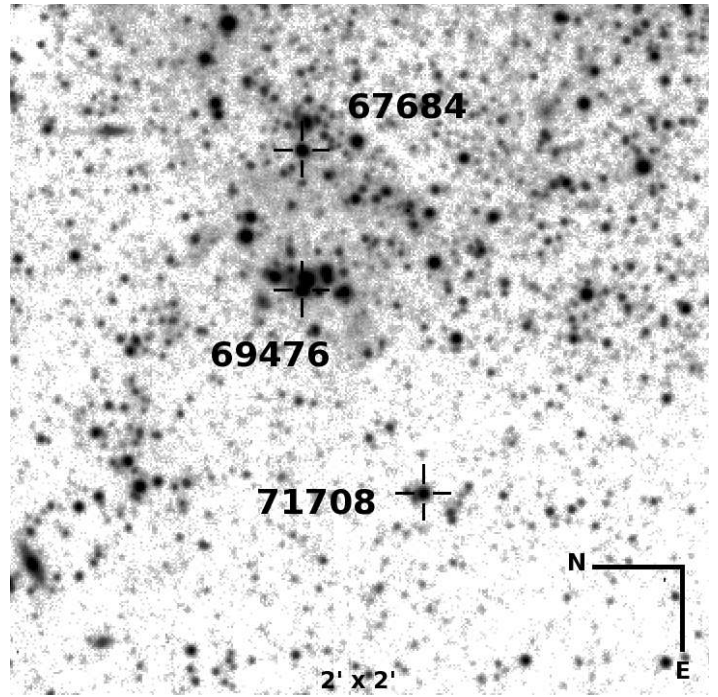}
   \caption{Stars 67684 (O8.5~I), 69476 (O6.5~III) and 71708 (late-O~III+neb).
           }
      \label{F:fc2}
\end{figure}

\begin{figure}
\centering
\includegraphics[width=0.50\textwidth]{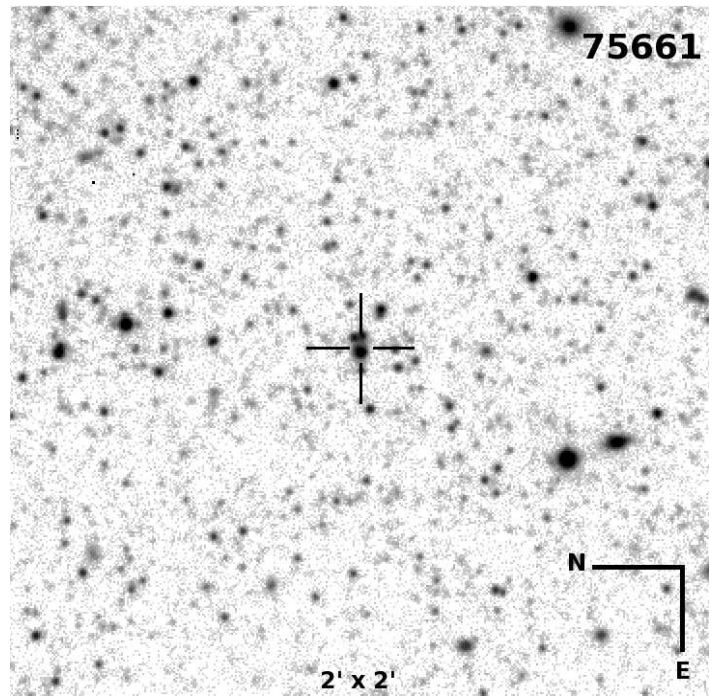}
   \caption{Star 75661 (O8~III).
           }
      \label{F:fc4}
\end{figure}

\begin{figure}
\centering
\includegraphics[width=0.50\textwidth]{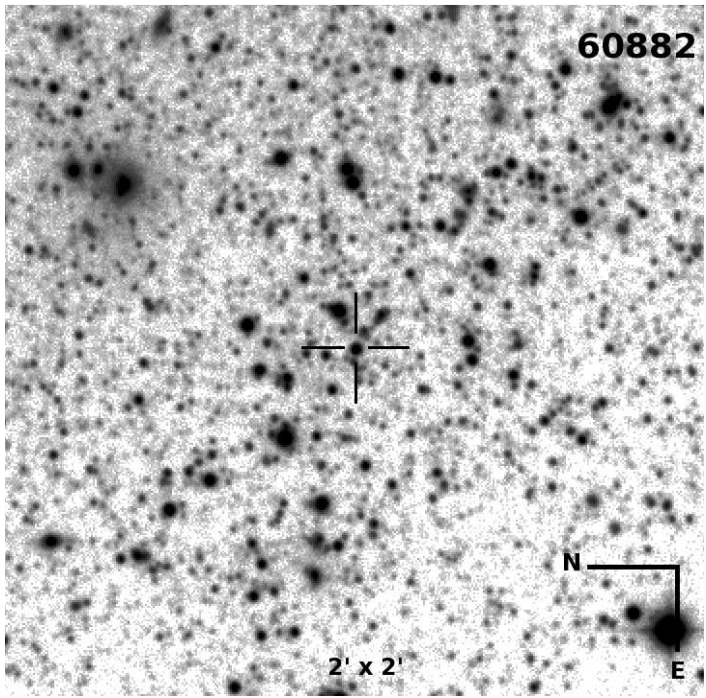}
   \caption{Star 60882 (B0.5~I-III).
           }
      \label{F:fc5}
\end{figure}

\begin{figure}
\centering
\includegraphics[width=0.50\textwidth]{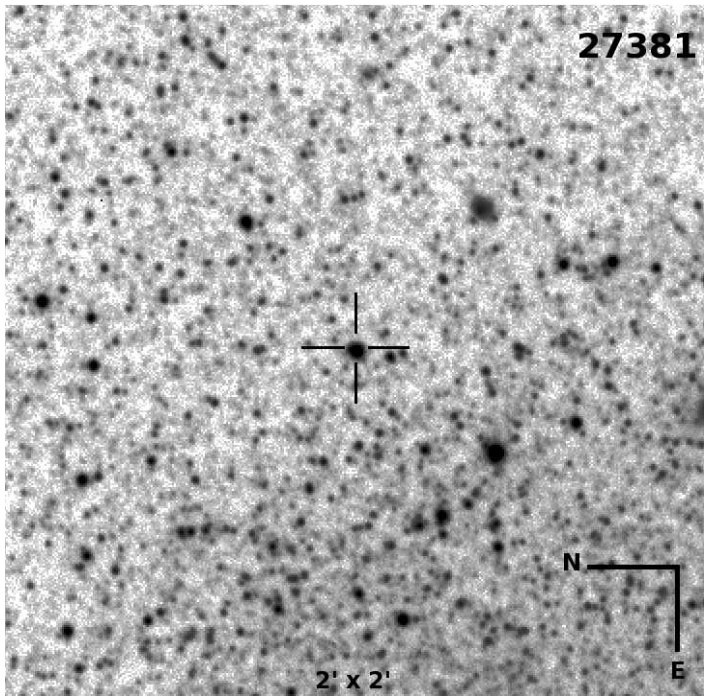}
   \caption{Star 27381 (B1-1.5~I).
           }
      \label{F:fc6}
\end{figure}

\begin{figure}
\centering
\includegraphics[width=0.50\textwidth]{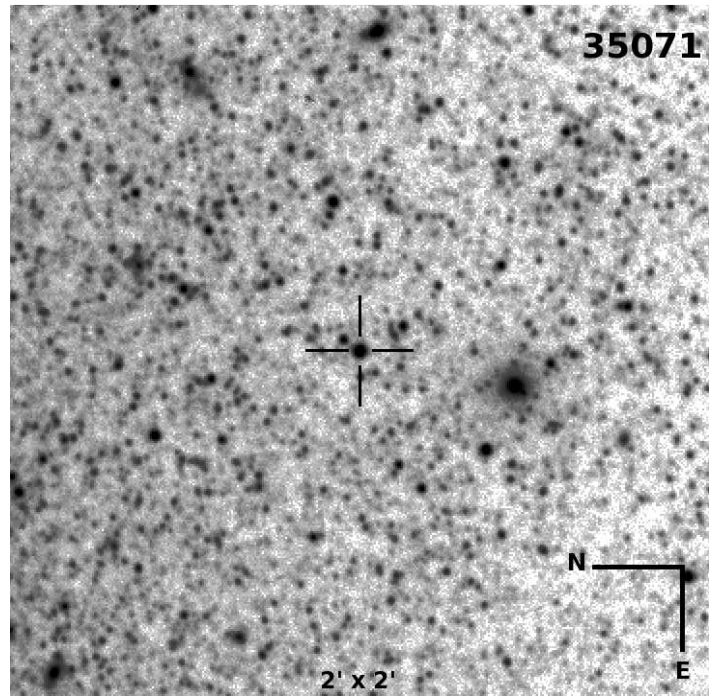}
   \caption{Star 35071 (B2.5~III).
           }
      \label{F:fc7}
\end{figure}

\end{appendix}

\end{document}